\documentclass[11pt]{article}
\usepackage{amsmath}
\usepackage{graphicx,psfrag,epsf}
\usepackage{enumerate}
\usepackage{natbib}
\usepackage{comment}
\usepackage{bm}
\usepackage{url} 
\usepackage[inline]{enumitem}

\newcommand{\independent}{\perp \!\!\! \perp}

\usepackage[margin=1in]{geometry}

\usepackage[colorlinks=true,allcolors=blue]{hyperref}
\usepackage[noabbrev,capitalize]{cleveref}

\usepackage{float}
\usepackage{tikz}
\usetikzlibrary{positioning}

\usepackage{xargs,xstring}

\usepackage{enumitem}

\begin{document}

\def\spacingset#1{\renewcommand{\baselinestretch}%
{#1}\small\normalsize} \spacingset{1}

  \title{\bf Discussion of the manuscript: \\ Spatial+: a novel approach to spatial confounding}
  \author{Georgia Papadogeorgou \\
   Department of Statistics, University of Florida}
	\date{}
  \maketitle

\bigskip


\spacingset{1} 

I congratulate Dupont, Wood and Augustin (DWA hereon) for providing an easy-to-implement method for estimation in the presence of spatial confounding, and for addressing some of the complicated aspects on the topic. The method regresses the covariate of interest on spatial basis functions and uses the residuals of this model in an outcome regression. The authors show that, if the covariate is not completely spatial, this approach leads to consistent estimation of the conditional association between the exposure and the outcome.

Below I discuss conceptual and operational issues that are fundamental to inference in spatial settings: (i) the target quantity and its interpretability, (ii) the non-spatial aspect of covariates and their relative spatial scales, and (iii) the impact of spatial smoothing. While DWA provide some insights on these issues, I believe that the audience might benefit from a deeper discussion. In what follows, I focus on the setting where a researcher is interested in interpreting the relationship between a given covariate and an outcome. I refer to the covariate of interest as the {\it exposure} to differentiate it from the rest.

\section{What are we estimating?}
\label{sec:target}

First, I discuss the target of estimation in the manuscript and in the literature. This target is key in deciding what is the appropriate method for estimation.

I begin by introducing some notation.
Assume that $Z$ and $Y$ are the exposure and outcome of interest, and covariates $\bm X$ include all the variables that confound this relationship. I use the subscript $i$ to denote the value of the variables at location $i = 1, 2, \dots, n$. In what follows, I assume that $\bm X$ is comprised of four random variables:
\begin{enumerate*}[label=(\alph*)]
	\item a measured variable $C$ that can be spatial or not, and unmeasured variables of the following three types:
	\item $S_1$ that is {\it completely spatial},
	\item $S_2^+$ that includes a spatial and an independent component, and
	\item $U$ that is not spatial.
\end{enumerate*}
Similarly to DWA, we can think of $S_2^+$ as a random variable that can be written as $S_2^+ = S_2 + E$ for $S_2$ being a completely spatial random variable and $E$ representing a component without spatial structure.

For simplicity of discussion, I also assume that the following additive model holds:
\begin{equation}
Y_i = \beta_0 + \beta_1 Z_i + \beta_2 C_i + \beta_3 S_{i1} + \beta_4 S_{i2}^+ + \beta_5 U_i + \epsilon_i,
\label{eq:model}
\end{equation}
where $\epsilon_i \overset{iid}{\sim} N(0, \sigma^2)$. This model implies that any spatial structure in the outcome $Y$ is due to spatial dependencies in its predictors, though this assumption could be relaxed.

\subsection{The ``unconditional'' association}
\label{subsec:RSR}

A researcher only has access to $(C_i, Z_i, Y_i)$, for $i = 1, 2, \dots, n$, while the remaining three random variables in $\bm X$ are not measured. With these data available, a first-step analysis could regress the outcome, $Y$ on the exposure and the sole observed covariate, $(Z, C)$. Even if not directly stated, the target of estimation in this regression is the relationship between the exposure and outcome of interest, while conditioning solely on the measured variable:
\begin{align}
(Y, Z) \mid C. \tag{Unconditional target}
\label{eq:unconditional_target}
\end{align}
Even though the remaining variables in $\bm X$ are not accounted for, this relationship can be of explicit scientific interest, especially if the measured variable $C$ is an interpretable characteristic of the population at hand. For example, $Z,$ $Y$ and $C$ can represent a subject's exposure to a given pollutant, their blood pressure levels, and age, respectively. In this case, the unconditional target represents the expected change in an individual's blood pressure for a unit increase in the pollutant, after adjusting for age. I refer to this target as the {\it unconditional} association as it does not condition on the spatial random variables. For {\it this} target of estimation, the OLS estimator will be {\it un}biased.

However, not conditioning on the remaining variables in $\bm X$ leads to complications for {\it inference} for the estimated coefficient. If at least one of the coefficients of the spatial predictors are non-zero ($\beta_3 \neq 0$, or $\beta_4 \neq 0$), the residuals of this regression would be spatially correlated. Since correlation in the model residuals that is not accounted for would lead to incorrect inference on model coefficients, researchers often include a spatial random effect with covariance structure that aims to mimic the residual dependence. The goal of the random effect is to explain the spatial dependence in the outcome that is {\it not} already explained by $(Z, C)$. However, if the exposure $Z$ is, at least partially, spatial, and since the additional model components are themselves spatial, they are often spuriously associated due to structural dependencies \citep{hodges2010adding, lee2020network}. Therefore, the random effect ``involuntarily'' captures part of the spatial variability in the outcome that was originally explained by the exposure, and the coefficient of the exposure in the model with the spatial random effect no longer has the same interpretation as the coefficient of the non-spatial model.

The restricted spatial regression (RSR) approach ensures that the random effect is orthogonal to the exposure of interest. That means that the target of estimation for RSR is the same as the one from the non-spatial model, and RSR is truly useful, applicable, and {\it un}biased for the relationship between the exposure and the outcome conditional {\it only} on the measured variables.

\subsection{The ``spatially-conditional'' association}
\label{subsec:spatial_target}

The target of estimation for DWA is different. Instead of focusing on the $(Y, Z)$ relationship when conditioning solely on $C$, DWA aim to also adjust for all spatial confounders. Informally, the target is set to learning
\begin{align}
(Y, Z) \mid C, \text{ all spatial confounders}, \tag{Spatially-conditional target}
\label{eq:spatial_target}
\end{align}
though as we will see below, it truely achieves an alternative quantity.
Therefore, in the presence of unmeasured spatial confounders, the targets of estimation for RSR and Spatial+ are different, the coefficient of the exposure in the two models has a different interpretation, and the methods should not be directly compared.

Since confounders are predictors of the outcome, an outcome model that does not adjust for spatial confounders will exhibit residual spatial dependence. At the same time, since they are predictors of the exposure, adding components in the regression model that are orthogonal to the exposure (as in RSR) will not capture them. However, since conditioning on any spatial variable that is {\it not} a confounder does not change the relationship under study, the spatially-conditional target above can be re-written as
\[
(Y, Z) \mid C, \text{ all spatial variables}.
\]

Writing the target of estimation explicitly as conditioning on any spatial variable clarifies two aspects of the manuscript.
\begin{enumerate*}[label=(\arabic*)]
\item If the exposure $Z$ is fully spatial and it does not include any independent components, then the spatially-conditional relationship cannot be learnt without additional assumptions, since the exposure is perfectly collinear with the conditioning set.~\footnote{To make progress in this setting, one might need to make assumptions on the relative spatial scales of the spatial confounders and the exposure, though I ignore this scenario here. This setting is investigated in \cite{schnell2020mitigating}, and a related discussion in terms of the spatial scales of the spatial confounders themselves can be found in \cref{subsec:assumptions}.}

\item If $Z$ includes an independent component, the spatially-conditional relationship between the exposure and the outcome can be learnt while focusing on the {\it non}-spatial aspect of the data.
\end{enumerate*}
The former is central in DWA's manuscript. The latter motivates three of the approaches discussed in the manuscript: the spatial model that includes basis functions directly in the outcome model, the gSEM model that controls for space both in the exposure and in the outcome model, and the proposed Spatial+ model that regresses the exposure itself on the basis functions and uses the exposure model residuals in the outcome model. When no spatial smoothing is performed, all of these methods adjust for any completely spatial structure in the data, and are therefore unbiased for the spatially-conditional association, as expected, and as illustrated in the simulations. The issues that arise with smoothing are discussed in \cref{sec:smoothing}.

The latter observation is also key in \cite{schnell2020mitigating}, where the authors link this observation to the assumption of positivity in causal inference, and discuss that estimation of the exposure coefficient can only be accurately achieved if the exposure itself varies within levels of the spatial confounder. In the setting of DWA where the spatial confounder can vary at {\it any} spatial frequency, this observation by \cite{schnell2020mitigating} is translated to $Z$ including an independent component. 

As discussed in \cref{subsec:RSR}, RSR studies the exposure-outcome relationship when conditioning only on the measured variables. As discussed above, Spatial+ aims to extend the conditioning set to include spatial confounders.
Even though conditioning on a more complete set of confounders is generally preferred, I illustrate that the estimated quantities from Spatial+ are not always interpretable. Take for example the true outcome model in \cref{eq:model}, and assume that both spatial variables are important $(\beta_3 \neq 0, \beta_4 \neq 0)$ while the independent variable is not ($\beta_5 = 0$). Then, we can re-write this model as
\[
Y_i = \beta_0 + \beta_1 Z_i + \beta_2 C_i + \beta_3 S_{i1} + \beta_4 (S_{i2} + E_i) + \epsilon_i,
\]
where $S_2$ and $E$ decompose $S_2^+$ in its spatial and independent components. The target of estimation that conditions on measured variables and {\it all} spatial confounders would be
\[
(Y, Z) \mid C, S_1, S_2^+ \quad \equiv \quad (Y, Z) \mid C, S_1, S_2, E.
\]
Therefore, this target would represent the exposure-outcome relationship when conditioning on the variables $C, S_1, S_2^+$.
However, the $E$ part of $S_2^+$ is {\it not} spatial and cannot be captured by any spatial basis function. Therefore, even though the implicit target would condition on {\it all} spatial variables, the method instead estimates
\begin{equation}
(Y, Z) \mid C, S_1, S_2.
\tag{Spatially-conditional quantity achieved}
\label{eq:spatial_achieved}
\end{equation}
Since the achieved target conditions solely on {\it part} of a confounder, the estimated conditional exposure-outcome relationship has a complicated interpretation. Whether this relationship is more or less interpretable than the unconditional target of \cref{subsec:RSR} is, I believe, up for debate!

In the previous blood pressure example, let's think of $S_1$ as representing outdoor temperature which is completely spatial, and $S_2^+$ as representing eating habits (such as intake of salt, or fried food). Even though eating habits are partially spatial due to local culture, they also include an independent, subject-specific component. Then, the coefficient estimated by Spatial+ would describe the expected change in blood pressure for a unit increase in the pollutant when adjusting for age, temperature, and {\it part} of someone's eating habits.

\subsection{Assumptions for interpretable spatially-conditional target}
\label{subsec:assumptions}

Therefore, the achieved target of estimation for the Spatial+ model often has a complicated interpretation, largely due to variables that include both spatial and independent components. So when can Spatial+ be used to acquire interpretable conditional associations?

The first case corresponds to the scenario where there are no spatial confounders of the exposure-outcome relationship that include an independent component \citep{papadogeorgou2019adjusting}. This assumption is implicitly made in equation (1) of DWA. If all covariates in the analysis are either fully spatial or not spatial at all (if, for example, $S_2^+ = S_2$), then Spatial+ estimates the exposure-outcome relationship conditional on all spatial variables, which has a clear interpretation. In turn, if there exist covariates that include both spatial and independent components such as $S_2^+$, these variables must not confound the exposure-outcome relationship when conditioning on fully spatial variables. This can be satisfied if
\[
E \independent Y \mid Z, C, S_1, S_2, \quad \text{or} \quad E \independent Z \mid C, S_1, S_2
\]
where $\independent$ is used to denote independence.
However, since all of these spatial variables are unmeasured, assuming that none of the spatial confounders includes an independent component remains, exactly that, {\it an assumption}.

So what happens when we are not willing to make the assumption that all spatial confounders are fully spatial? Progress can still be made in this setting if we are willing to make an alternative assumption on the relative spatial scales of spatial confounders, and update Spatial+ accordingly. Remember that the issue arises because $S_1$ and $S_2^+$ are both spatial and unmeasured, which forbids us from directly differentiating between the two. To make progress, we could make an assumption which allows us to separate the one spatial variable from the other.
Such an assumption could be the following: {\it $S_2^+$ varies only in higher spatial frequencies than $S_1$.}
If this assumption holds, then including spatial basis functions that represent only low-frequency spatial components would capture $S_1$ fully, while not capturing any aspect (spatial or independent) of $S_2^+$ \citep{antonelli2017spatial, keller2020selecting}. This could be acheived, for example, using wavelet or fourier basis functions for which the frequency of the basis functions used is directly controllable. Under this assumption and the corresponding adjustment of Spatial+, the method would be estimating the exposure-outcome relationship when conditioning on $C$ and $S_1$ only, which would have a clear interpretation.

\subsection{Spatial confounding when estimating causal effects}

If the set of covariates $\bm X$ includes all the confounders of the exposure-outcome relationship, then an outcome model which adjusts for the exposure and all of the confounders and is correctly specified can be the base for estimating causal effects \citep[][g-computation]{robins1986new, snowden2011implementation}. I focus first on linear additive models such as the one in \cref{eq:model}. In this case, the coefficient of the exposure, $\beta_1$, can be interpreted causally as the effect of increasing the exposure $Z$ by one unit on the expected outcome $Y$. However, the coefficient will not have the same, causal interpretation if we do not adjust for the variables $S_1,$ $S_2+$ and $U$, which are unmeasured, nor if we only adjust for a subset of them. Since $U$ is a completely independent random variable, we cannot hope to identify the causal effect of the exposure on the outcome in this manner. What's more, adjusting for {\it some} of the confounders, and not all, does {\it not} necessarily mean that the bias for estimating causal effects is reduced. Instead, it could {\it increase}, if confounding from spatial and independent variables are in opposite directions! Therefore, if unmeasured independent variables confound the exposure-outcome relationship, adjusting for spatial unmeasured confounders does not necessarily bring us ``closer'' to estimating causal effects.

However, if the true outcome model is non-linear, or includes exposure-covariate interactions, the complications are not resolved, even if we assume the absence of independent, unmeasured confounders (for example, if $U$ does not exist, and $S_2+$ is fully spatial).
In order to use the outcome model for estimating causal effects, g-computation teaches us that we need to integrate over the distribution of the confounders, and the coefficient of the exposure is not necessarily causally interpretable (though log-linear models are an exception for a specific type of causal contrast \citep{schnell2020mitigating}). When the spatial confounders are unmeasured, and their distribution cannot be learnt from data, integrating over their distribution cannot be achieved directly, and requires additional assumptions \citep[see, for example][]{christiansen2020towards}. 
Recently, \cite{reich2021review} wrote a review article regarding spatial causal inference, including a thorough discussion on spatial confounding.

\section{Spatial smoothing}
\label{sec:smoothing}

Based on the discussion of \cref{sec:target}, we have established that the non-spatial and RSR models estimate a different quantity compared to the rest. We have also established that regressing away all spatial aspect of the outcome or the exposure (through, for example, {\it unpenalized} splines) would adjust fully for the spatial component of all spatial confounders. Assuming that this quantity has a useful interpretation, here I focus on investigating the following questions: {\it what happens with spatial smoothing}, and {\it how much should predictive accuracy matter?}

Since spatial smoothing excludes the higher frequency components, it implies that the achieved target represents the exposure-outcome association conditional on the spatial variables that vary at the included, lower frequencies {\it only}. If all spatial confounders vary at lower frequencies, the estimated quantity will truly align with the spatially-conditional association. However, even though spatial smoothing is expected to lead to lower prediction MSE, it is not immidiately clear whether it should be used for coefficient estimation. In the presence of high frequency spatial confounders, spatial smoothing might lead us to ``miss'' these confounders.

I investigate two scenarios where the Spatial and Spatial+ models might miss strong, high frequency spatial confounders, and return biased estimates of the spatially-conditional association.~\footnote{Remember that the Spatial model employs spatial basis functions in the outcome model.}
These spatial variables can be strong confounders in at least two different ways:
\begin{enumerate*}[label=(\arabic*)]
\item they can be important predictors of the exposure and weak predictors of the outcome, or
\item they can be weak predictors of the exposure, and important predictors of the outcome.
\end{enumerate*}
In the former case, since the high-frequency spatial variables are strong predictors of the exposure, one hopes that the choice of penalty in the exposure model for Spatial+ will not penalize these high frequency components heavily. The reverse is true for the Spatial model, for which we would expect that it will penalize the high frequency spatial basis functions more heavily, as these are weak predictors of the outcome. Therefore, in this scenario, we would expect Spatial+ to perform better than the Spatial model for estimating the spatially-conditional association.
The reverse is true in the latter case. If the spatial confounders are weak predictors of the exposure, it is expected that the chosen penalty for the exposure model according to Spatial+ will penalize the high frequency components heavily, returning biased estimates of the spatially-conditional association. In contrast, we would expect the Spatial model that includes spatial basis functions in the outcome model directly to perform more accurately.

Even though it is prudent not to overly generalize the message from this discussion, I believe that deeper investigation is warranted on how the methods perform under different confounding scenarios.
That is essential under the light of similar observations from other parts of the literature that focus on estimation rather than prediction, and encourage the use of {\it both} an exposure and an outcome model simultaneously \citep[e.g.,][]{wang2012bayesian, belloni2014inference, shortreed2017outcome, antonelli2019high}. Specificially in the setting of spatial confounding, the use of both models is proposed in at least two approaches (gSEM, investigated in the manuscript; \cite{thaden2018structural} and \cite{schnell2020mitigating}).
If we view gSEM as a combination of the Spatial and Spatial+ models in that it adjusts for spatial effects in both the exposure and the outcome model, one might expect gSEM to be more robust to high frequency spatial confounding than any of the two methods separately.

Finally, the work referenced above has been crucial in illuminating that {\it prediction} and {\it estimation} are two separate goals, and that methodology that is appropriate for one can be inappropriate for the other. In light of this work,
even though the methods' predictive accuracy can inform us of how well each model fits the data,
it should {\it not} be used as a model-choice criterion when we are interested in interpreting estimated parameters. In fact, a model with lower AIC might often have higher bias for the coefficient of interest.

\bibliographystyle{agsm}
\bibliography{Biometrics_discussion}

\end{document}